Growth of $RE$Ba$_2$Cu$_3$O$_x$ single-crystal whiskers utilizing the concept of high-entropy alloys

Yasunori Suzuki, Masanori Nagao*, Yuki Maruyama, Satoshi Watauchi, and Isao Tanaka

*University of Yamanashi, 7-32 Miyamae, Kofu, Yamanashi 400-0021, Japan*

E-mail: mnagao@yamanashi.ac.jp

Single-crystal whiskers of $RE$Ba$_2$Cu$_3$O$_x$ ($RE$-123, $RE$: rare earth element) cuprate superconductors were successfully grown using Te-doped precursors utilizing the concept of high-entropy-alloys (HEAs) at the $RE$ site. The obtained whiskers were 0.5–1.0 mm length with approximately 10 μm thickness. The width of the flat surface, which corresponds to the $ab$-plane, was 10–50 μm. The $RE$ site of the grown whiskers could be easily substituted by large ionic radius $RE$ elements (Gd, Dy, and Ho). Substitution with small ionic radius $RE$ elements (Tm, Yb, and Lu) resulted in the opposite trend. The superconducting transition temperature and superconducting anisotropy of the grown $RE$-123 whiskers were 89–93 K and 6–10, respectively. The mixed entropy at the $RE$ site in the $RE$-123 whiskers did not affect the superconducting transition temperature and superconducting anisotropy.

## 1. Introduction

High-entropy alloys (HEAs) are defined as alloys containing more than five elements with concentrations between 5 and 35 at%. HEAs exhibit high hardness and excellent resistance[1,2] and have been investigated as functional materials.[3,4] Some of them also exhibit excellent irradiation resistance,[5-7] indicating that HEAs are suitable for a high-energy particle irradiation environment such as that inside a nuclear fusion reactor. Recently, the concept of HEA was introduced in superconducting materials, and the robustness of the superconducting states under extreme conditions[8] and the improvement in superconducting properties were investigated.[9-11] Studying the growth of the single crystals of such superconducting materials has been crucial for understanding the intrinsic properties of these materials.[12-15]

$RE$Ba$_2$Cu$_3$O$_x$ ($RE$-123, $RE$: rare earth elements) cuprate superconductors[16] are expected to be employed in superconducting application such as the generation of a magnetic field at liquid nitrogen temperature (77 K) because the superconducting transition temperature ($T_c$) is higher than 77 K. Thin films and bulk samples of polycrystalline $RE$-123 have been prepared using the concept of HEAs at the $RE$ site.[11,17-19] Furthermore, the superconducting critical current can be increased by introducing the concept of HEAs at the $RE$ site.[20] Although magnets made with $RE$-123 superconductors can be used for plasma containment in a nuclear fusion reactor,[21] their exposure to the high-energy neutron irradiation environment in these reactors damages them. As previously mentioned, the concept of HEA can be utilized to achieve enhanced irradiation resistance.[5-7] Therefore, *RE*-123 prepared utilizing the concept of HEA can be employed for magnet application in a nuclear fusion reactor. *RE*-123 single crystals with the HEAs concept affecting the irradiation property are meaningful investigation targets. However, a systematic investigation of the superconductivity and the high-entropy effect at the *RE* site in single crystals has not yet been performed.

We focus on *RE*-123 prepared utilizing the concept of HEA at the *RE* site of single-crystal whiskers. Such single-crystal whiskers are suitable for investigating the intrinsic properties because of their useful characteristics such as perfect crystallinity, predictable growth directions, and excellent superconducting properties. Previously, we developed a whisker growth method using Te-doped precursors[22-24] and succeeded in growing *RE*-123 single-crystal whiskers substituted with various single kinds of *RE* elements.[25] Applying this method, we grew *RE*-123 single-crystal whiskers utilizing the concept of HEAs at the *RE* site.

In this paper, we report the synthesis of *RE*-123 single-crystal whiskers grown utilizing the concept of HEA at the *RE* site using Te-doped precursors. Subsequently, the superconducting properties and the high-entropy effect at the *RE* site were investigated in terms of the superconducting transition temperature and superconducting anisotropy. The superconducting properties of the *RE*-123 single-crystal whiskers were affected by Ba-site substitution, which depends on the ionic radius of the *RE* elements.[25] Thus, Ba-site substitution should be suppressed for investigating the high-entropy effect at the *RE* site. Consequently, Dy, Ho, Er, Tm, Yb, and Lu, which could be barely substituted at the Ba site, were chosen as the *RE* elements. Moreover, because Gd has a lower propensity for Ba-site substitution compared to La-Eu, it was employed for achieving a higher *RE*-site entropy.

## 2. Experimental methods

$REBa_2Cu_3O_x$ superconducting single-crystal whiskers were grown from Te-doped precursors utilizing the concept of HEAs at the *RE* site.[22-25] A precursor powder with a nominal composition of $RE_{1.5}Ba_3Cu_3Te_{0.5}O_x$, with $RE = Gd_aDy_bHo_cEr_dTm_eYb_fLu_g$ ($a+b+c+d+e+f+g = 1.5$), was synthesized through a solid-state reaction. The starting materials, $RE_2O_3$, $BaCO_3$, and CuO, and the flux, $TeO_2$, were ground completely in a mortar and calcined in air at 750–800 °C for 10 h. The grinding and calcination of the reactants were repeated three times. The calcined powder (0.5 g) was pressed into a pellet of 10 mm diameter and approximately 3 mm thickness. The precursor pellets were heated in dry air at $T_{max}$ for 10 h, followed by slow cooling to 900 °C at a rate of 0.5 °C/h. Then, the temperature of the furnace controller was set to cool to room temperature (30 °C) at 8 h. However, the actual furnace temperature did not become the set temperature below 300 °C. Therefore, temperatures below 300 °C resulted in spontaneous cooling. The $T_{max}$ values were varied between 940 and 990 °C and optimized for the growth length of the whiskers.

Scanning electron microscopy (SEM) was conducted using a TM3030 system from Hitachi High Technologies, Japan. The compositional ratios of the *RE*-123 whiskers were evaluated using energy-dispersive X-ray spectroscopy (EDS; Quantax 70, Bruker Corp., Germany). The analytical content of each element was defined as $C_{RE}$ (*RE*: Gd, Dy, Ho, Er, Tm, Yb, and Lu). The obtained values were normalized using the atomic concentration using the relation $\Sigma(C_{RE}) = 1.00$. Subsequently, Ba and Cu contents ($C_{Ba}$ and $C_{Cu}$, respectively) were estimated to the precision of two decimal places. X-ray diffraction (XRD, MultiFlex, Rigaku, Japan) with $2\theta/\theta$ scan using Cu-K$\alpha$ radiation was performed to confirm the crystal structure. The mixed entropy ($\Delta S_{mix}$) at the *RE* site was calculated from the analytical contents using the following equation:[1]

$$\Delta S_{mix} = -R \Sigma (C_{RE} \ln C_{RE}) \quad (RE: Gd, Dy, Ho, Er, Tm, Yb, Lu) \quad \text{(Eq. 1)}$$

Here, $R$ = 8.314 J/mol·K (gas constant). In an alternative definition, HEAs are defined as compounds with $\Delta S_{mix}$ values greater than $1.5R$.[26] This definition has been used in this study.

The transport properties of the whiskers were determined by the standard four-probe method in the constant-current density ($J$) mode using a physical property measurement system (PPMS DynaCool, Quantum Design, USA). The obtained whiskers were placed on a MgO single-crystal substrate, and the electrical terminals were fabricated using a Ag paste (H20E, EPOXY TECHNOLOGY, USA) that was annealed at 450 °C for 10 min in air.

The criterion of superconducting transition temperature with zero resistivity ($T_c^{zero}$) was defined as 10 μΩcm in the resistivity–temperature ($\rho$–$T$) characteristics. The superconducting anisotropy ($\gamma_s$) was estimated by measuring the angular ($\theta_a$) dependence of resistivity ($\rho$) in the liquid-state flux under various magnetic fields ($H$) and applying an effective mass model.[27] The relationship between mixed entropy ($\Delta S_{mix}$) and superconducting properties ($T_c^{zero}$ and $\gamma_s$) was established.

## 3. Results and discussion

We successfully grew *RE*-123 cuprate single-crystal whiskers substituted with 2–7 *RE*

elements (using the concept of HEAs) at the *RE* site. Figure 1 shows a typical SEM image of the grown *RE*-123 whisker. The length and thickness of the grown whiskers were 0.5–1.0 mm and approximately 10 μm, respectively. The whiskers possessed a flat surface that was 10–50 μm wide. Evidently, the width of the flat surface was larger than the thickness. The characteristics of the grown whiskers are summarized in Table I. XRD measurement of the whiskers was performed on the flat surface. Figure 2 shows a typical XRD pattern of the whisker (Sample RE7; Table I). The pattern shows only the 00*l* diffraction peak indices of the *RE*-123 structure. Therefore, the single-crystal whiskers were confirmed to possess the typical *RE*-123 structure. The flat surface was the *ab*-plane, as estimated from the XRD patterns. The full width at half maximum (FWHM) of the single-crystal whiskers was determined to be 0.07°–0.11° from the (006) diffraction peak. *RE*-123 single-crystal whiskers with various *RE* contents were grown from the nominal compositions (*a-g*). Figure 3 shows the relationship between the nominal composition and averaged analytical content of each *RE* element, which are normalized using the relations $a+b+c+d+e+f+g = 1.0$ and $\Sigma(C_{RE}) = 1.00$, respectively. For *RE* elements with large ionic radii (Gd, Dy, and Ho), the analytical contents were higher than those of the nominal compositions. On the other hand, the opposite trend was observed for *RE* elements with small ionic radii (Tm, Yb, and Lu). This indicates the *RE* elements with large ionic radii are easily substituted in the *RE*-123 whiskers. Superconducting properties were measured for seven kinds of *RE*-123 whiskers for each number of *RE* elements in all the obtained *RE*-123 whiskers. Analytical contents, heat treatment temperature ($T_{max}$), *c*-axis lattice constants, average ionic radii of *RE* sites, and $\Delta S_{mix}$ at the *RE* site of these *RE*-123 whiskers are summarized in Table I. The superconducting properties of *RE*-123 superconductors are affected by the average ionic radius of the *RE* sites. Consequently, whiskers with a similar average ionic radius of the *RE* sites were selected because the effect of $\Delta S_{mix}$ at the *RE* site was being investigated. The *c*-axis lattice constants at the various *RE* site elements were similar and estimated to be in the range of 11.63–11.66 Å. The average compositional values of Ba- and Cu normalized using the relation $\Sigma(C_{RE}) = 1.00$ for the selected whiskers were 2.1±0.1 and 3.1±0.2, respectively. Accordingly, substitution of the Ba site by *RE* elements was expected to be low. On the other hand, Te from the flux was not detected in the grown single-crystal whiskers with a minimum sensitivity limit of approximately 1 wt%.

Figure 4 (a) shows the $\rho$–$T$ characteristics of the *RE*-123 single-crystal whiskers along the *ab*-plane in the temperature range of 60–300 K. The superconducting transition with zero resistivity ($T_c^{zero}$) was observed in the 88.8–92.7 K range, and metallic behavior was observed in the normal state (higher than 100 K). The dashed lines for each sample in Fig. 4 are linear extrapolations of the resistivity from the normal state. Figure 4 (b) shows that the slopes of the $\rho$–$T$ plots in the normal state are uncorrelated to $\Delta S_{mix}$.

Figure 5 shows the temperature dependence of the resistivity along the $ab$-plane in the temperature range of 100–70 K under 0.1–9.0 T of magnetic field ($H$) parallel to the (a) $ab$-plane and (b) $c$-axis for the as-grown RE5 sample. The suppression of the critical temperature under the magnetic field applied parallel to the $c$-axis was more than that applied parallel to the $ab$-plane, and a typical $RE$-123 superconducting behavior was observed. The other $RE$-123 whiskers also exhibited a similar behavior.

The resistivity ($\rho$) was measured under different magnetic fields ($H$) in the flux liquid state to estimate the superconducting anisotropy ($\gamma_s$) using previously reported methods.[28,29] The reduced field ($H_{red}$) was calculated using the following equation for an effective mass model:

$$H_{red} = H(\sin^2\theta_a + \gamma_s^{-2}\cos^2\theta_a)^{1/2} \quad \text{(Eq. 2)}$$

Here, $\theta_a$ is the angle between the $ab$-plane and the magnetic field.[27] $\gamma_s$ was estimated from the best scaling of the $\rho$–$H_{red}$ relationship. Figure 6 shows the $\theta_a$ dependence of $\rho$ for different magnetic fields ($H$ = 0.1–9.0 T) in the flux liquid state ($T/T_c^{zero}$ = 0.99) for the RE5 single-crystal whisker. The $\rho$–$\theta_a$ curve exhibited a bilateral symmetry at $\theta_a$ = 90. Small dips were observed in the $\rho$–$\theta_a$ curves around the $H$ // $c$-axis at magnetic fields less than 1.9 T. A similar behavior was observed in the other obtained whiskers. The small dip originated from twin boundaries in the whiskers, as was reported for a Y-123 single crystal with twin boundaries.[30] The twin boundaries were observed on the *ab*-plane in the obtained whiskers using a polarization microscope. Figure 7 shows the $\rho$–$H_{red}$ scaling obtained from the $\rho$–$\theta_a$ curves in Figure 6 using Eq. 2. The scaling was performed by setting $\gamma_s$ = 6.5. The $T_c^{zero}$ and $\gamma_s$ values of the other *RE*-123 whiskers listed in Table I were also evaluated by the same method. The values of these superconducting parameters are summarized in Table II.

Figure 8 shows the dependence of $T_c^{zero}$ and $\gamma_s$ on the $\Delta S_{mix}$ at the *RE* site for the RE1–RE7 samples. There was no clear correlation of $T_c^{zero}$ or $\gamma_s$ with $\Delta S_{mix}$. In other words, the $\Delta S_{mix}$ at the *RE*-site for the *RE*-123 superconductor may not affect the superconducting transition temperature and superconducting anisotropy, in contrast with that observed for the oxygen content in the CuO chain layer.[31] When the result of Fig. 4 (b) is also made an addition, we expected that both superconducting properties and normal state transport properties were insensitive to the $\Delta S_{mix}$ which affects the local disorder.

## 4. Conclusions

*RE*-123 cuprate single-crystal whiskers substituted with 2–7 *RE* elements were successfully grown using Te-doped precursors. The mixed entropy values at the *RE* site ranged from 0.67*R* to 1.91*R*, with some of the compounds satisfying the concept of HEAs at the *RE* site. The *RE* elements with larger ionic radii were selectively substituted at the *RE* site of the grown whiskers. The superconducting transition temperatures of the *RE*-123 whiskers with a similar average ionic radius of the *RE* sites exhibited no dependence on the mixed entropy at the *RE* site. The superconducting anisotropy also did not depend on the mixed entropy at the *RE* site. These results suggested that the advantages of high entropy alloys may be appended without the reduction of superconducting properties for *RE*-123 superconductors.

**Acknowledgments**

This work was partially supported by JSPS KAKENHI (Grant-in-Aid for Scientific Research (B): Grant Number 21H02022 and Grant-in-Aid for Challenging Exploratory Research: Grant Number 21K18834). We would like to thank Editage (www.editage.com) for English language editing.

# References

1) J. W. Yeh, S. K. Chen, S. J. Lin, J. Y. Gan, T. S. Chin, T. T. Shun, C. H. Tsau, and S. Y. Chang, Adv. Eng. Mater. **6**, 299–303 (2004).

2) B. Cantor, I. T. H. Chang, P. Knight, and A. J. B. Vincent, Mater. Sci. Eng. A, **375–377**, 213–218 (2004).

3) O. N. Senkov, G. B. Wilks, J. M. Scott, and D. B. Miracle, Intermetallics **19**, 698–706 (2011).

4) Y. Yao, Z. Huang, P. Xie, S. D. Lacey, R. J. Jacob, H. Xie, F. Chen, A. Nie, T. Pu, M. Rehwoldt, D. Yu, M. R. Zachariah, C. Wang, R. Shahbazian-Yassar, J. Li, and L. Hu, Science **359**, 1489–1494 (2018).

5) Q. Xu, H. Q. Guan, Z. H. Zhong, S. S. Huang, and J. J. Zhao, Sci. Rep. **11**, 608 (2021).

6) K. Jin, C. Lu, L. M. Wang, J. Qu, W. J. Weber, Y. Zhang, and H. Bei, Scr. Mater. **119**, 65–70 (2016).

7) Y. Lu, H. Huang, X. Gao, C. Ren, J. Gao, H. Zhang, S. Zheng, Q. Jin, Y. Zhao, C. Lu, T. Wang, and T. Li, J. Mater. Sci. Technol. **35**, 369–373 (2019).

8) J. Guo, H. Wang, F. von Rohr, Z. Wang, S. Cai, Y. Zhou, K. Yang, A. Li, S. Jiang, Q. Wu, R. J. Cava, and L. Sun, Proc. Natl. Acad. Sci. USA. **114**, 13144–13147 (2017).

9) R. Sogabe, Y. Goto, and Y. Mizuguchi, Appl. Phys. Express **11**, 053102 (2018).

10) R. Sogabe, Y. Goto, T. Abe, C. Moriyoshi, Y. Kuroiwa, A. Miura, K. Tadanaga, and Y. Mizuguchi, Solid State Commun. **295**, 43–49 (2019).

11) Y. Shukunami, A. Yamashita, Y. Goto, and Y. Mizuguchi, Physica C **572**, 1353623 (2020).

12) Y. Fujita, K. Kinami, Y. Hanada, M. Nagao, A. Miura, S. Hirai, Y. Maruyama, S. Watauchi, Y. Takano, and I. Tanaka, ACS Omega **5**, 16819–16825 (2020).

13) T. Ying, T. Yu, Y. Shiah, C. Li, J. Li, Y. Qi, and H. Hosono, J. Am. Chem. Soc., **143**, 7042–7049 (2021).

14) L. A. Pressley, A. Torrejon, W. A. Phelan, and T. M. McQueen, Inorg. Chem., **59**, 17251–17258 (2020).

15) C. K. Fedon, Q. Zheng, Q. Huang, E. S. Choi, J. Yan, H. Zhou, D. Mandrus, and V. Keppens, Phys. Rev. Mater., **4**, 104411 (2020).

16) M. K. Wu, J. R. Ashburn, C. J. Thorng, P. H. Hor, R. L. Meng, L. Gao, Z. J. Huang, Y. Q. Wang, and C. W. Chu, Phys. Rev. Lett. **58**, 908–910 (1987).

17) K. Wang, Q. Hou, A. Pal, H. Wu, J. Si, J. Chen, S. Yu, Y. Chen, W. Lv, J.-Y. Ge, S. Cao, J. Zhang, and Zh. Feng, J. Supercond. Nov. Magn. **34**, 1379–1385 (2021).

18) A. Yamashita, K. Hashimoto, S. Suzuki, Y. Nakanishi, Y. Miyata, T. Maeda, and Y.

Mizuguchi, Jpn. J. Appl. Phys. **61**, 050905 (2022).

19) J. Chen, R. Huang, X. Zhou, D. Zhou, M. Li, C. Bai, Z. Liu, and C. Cai, J. Rare Earths (2022) in press.

20) A. Yamashita, Y. Shukunami, and Y. Mizuguchi, R. Soc. Open Sci. **9**, 211874 (2022).

21) I. Teruo and Y. Nagato, J. Plasma Fusion Res. **93**, 222–229 (2017).

22) M. Nagao, M. Sato, H. Maeda, S.-J. Kim, and T. Yamashita, Appl. Phys. Lett. **79**, 2612–2614 (2001).

23) M. Nagao, M. Sato, H. Maeda, S.-J. Kim, and T. Yamashita, Jpn. J. Appl. Phys. **41**, L43–L45 (2002).

24) M. Nagao, T. Kawae, K. S. Yun, H. Wang, Y. Takano, T. Hatano, T. Yamashita, M. Tachiki, and H. Maeda, J. Appl. Phys. **98**, 073903 (2005).

25) M. Nagao, S. Watauchi, I. Tanaka, T. Okutsu, Y. Takano, T. Hatano, and H. Maeda, Jpn. J. Appl. Phys. **49**, 033101 (2010).

26) L. Sun and R. J. Cava, Phys. Rev. Materials. **3**, 090301 (2019).

27) G. Blatter, V. B. Geshkenbein, and A. I. Larkin, Phys. Rev. Lett., **68**, 875–878 (1992).

28) Y. Iye, I. Oguro, T. Tamegai, W. R. Datars, N. Motohira, and K. Kitazawa, Physica

C **199**, 154–160 (1992).

29) H. Iwasaki, O. Taniguchi, S. Kenmochi, and N. Kobayashi, Physica C **244**, 71–77 (1995).

30) G. W. Crabtree, W. K. Kwok, U. Welp, J. Downey, S. Fleshler, K. G. Vandervoort, and J. Z. Liu, Physica C **185**–**189,** 282–287 (1991).

31) T. Katayama, J. Shimoyama, H. Ogino, S. Horii, and K. Kishio, J. Phys.: Conf. Ser. **97,** 12144 (2008).

**Figure captions**

**Figure 1.** SEM image of the *RE*-123 single-crystal whiskers prepared utilizing the concept of HEAs at the *RE* site.

**Figure 2.** Typical XRD pattern of the grown *RE*-123 single-crystal whisker (sample RE7).

**Figure 3.** Relationship between nominal compositions and averaged analytical *RE* contents normalized using the relations $a+b+c+d+e+f+g = 1.0$ and $\Sigma(C_{RE}) = 1.00$, respectively, for all the obtained *RE*-123 whiskers.

**Figure 4.** (a) $\rho$–$T$ characteristics along the *ab*-plane of the *RE*-123 single-crystal whiskers. The dashed lines are linear extrapolations of the resistivity from the normal state. (b) Slopes of the $\rho$–$T$ curves at the normal state as a function of $\Delta S_{mix}$.

**Figure 5.** $\rho$–$T$ characteristics along the *ab*-plane under 0.1–9.0 T of magnetic field (*H*) parallel to the (a) *ab*-plane and (b) *c*-axis for the as-grown RE5 sample.

**Figure 6.** Angular $\theta_a$ dependence of resistivity $\rho$ ($\rho$–$\theta_a$ curves) in flux liquid state ($T/T_c^{zero}$ = 0.99) at various magnetic fields (*H* = 0.1-9.0 T) for the as-grown RE5 sample.

**Figure 7.** Scaling of the $\rho$–$\theta_a$ curves at a reduced magnetic field $H_{red}$ for the as-grown RE5 sample using the data in Figure 6.

**Figure 8.** Dependence of (a) $T_c^{zero}$ and (b) $\gamma_s$ on $\Delta S_{mix}$ for the RE1-7 samples.

**Table I.** Nominal compositions, analytical contents, $T_{max}$, $c$-axis lattice constants, average ionic radius of $RE$ sites, and $\Delta S_{mix}$ at the $RE$ site in the $RE$-123 single-crystal whiskers

| Sample | Nom. Anal. | Gd:$a$ $C_{Gd}$ | Dy:$b$ $C_{Dy}$ | Ho:$c$ $C_{Ho}$ | Er:$d$ $C_{Er}$ | Tm:$e$ $C_{Tm}$ | Yb:$f$ $C_{Yb}$ | Lu:$g$ $C_{Lu}$ | Ba $C_{Ba}$ | Cu $C_{Cu}$ | $T_{max}$ (°C) | $c$-axis lattice constants (Å) | Average ionic radius of $RE$ sites (Å) | $\Delta S_{mix}$ ($R$ : Gas constant) |
|---|---|---|---|---|---|---|---|---|---|---|---|---|---|---|
| RE1 | Nom. |  |  |  | 1.5 |  |  |  | 3.0 | 3.0 | 965 | 11.65 | 1.00 | 0 |
|  | Anal. |  |  |  | 1.00 |  |  |  | 1.95(2) | 2.95(1) |  |  |  |  |
| RE2 | Nom. |  |  | 0.78 | 0.72 |  |  |  | 3.0 | 3.0 | 970 | 11.63 | 1.01 | 0.67$R$ |
|  | Anal. |  |  | 0.59(1) | 0.41(1) |  |  |  | 2.08(3) | 3.17(3) |  |  |  |  |
| RE3 | Nom. |  |  | 0.40 | 0.50 | 0.60 |  |  | 3.0 | 3.0 | 965 | 11.63 | 1.01 | 1.01$R$ |
|  | Anal. |  |  | 0.44(1) | 0.34(1) | 0.19(1) |  |  | 2.18(6) | 3.2(1) |  |  |  |  |
| RE4 | Nom. |  | 0.25 | 0.30 | 0.40 | 0.55 |  |  | 3.0 | 3.0 | 970 | 11.63 | 1.01 | 1.35$R$ |
|  | Anal. |  | 0.26(1) | 0.34(0) | 0.29(1) | 0.15(0) |  |  | 1.97(1) | 2.87(1) |  |  |  |  |
| RE5 | Nom. |  | 0.1 | 0.2 | 0.2 | 0.4 | 0.6 |  | 3.0 | 3.0 | 960 | 11.66 | 1.00 | 1.60$R$ |
|  | Anal. |  | 0.16(4) | 0.17(5) | 0.20(2) | 0.22(7) | 0.25(5) |  | 1.9(2) | 2.9(3) |  |  |  |  |
| RE6 | Nom. | 0.15 | 0.13 | 0.17 | 0.20 | 0.35 | 0.50 |  | 3.0 | 3.0 | 990 | 11.63 | 1.01 | 1.78$R$ |
|  | Anal. | 0.23(1) | 0.17(1) | 0.15(1) | 0.15(2) | 0.19(1) | 0.20(1) |  | 2.14(8) | 3.34(7) |  |  |  |  |
| RE7 | Nom. | 0.15 | 0.10 | 0.10 | 0.15 | 0.20 | 0.30 | 0.50 | 3.0 | 3.0 | 970 | 11.65 | 1.01 | 1.91$R$ |
|  | Anal. | 0.20(1) | 0.13(1) | 0.10(1) | 0.15(1) | 0.09(1) | 0.20(1) | 0.15(1) | 2.0(1) | 3.0(2) |  |  |  |  |

Nom.: Nominal composition ($a+b+c+d+e+f+g = 1.5$), Anal.: Analytical content (Normalized using $\Sigma(C_{RE}) = 1.00$)

**Table II.** Evaluated $T_c^{zero}$ and $\gamma_s$ values for the *RE*-123 single-crystal whiskers

| Sample | Superconducting properties | |
|---|---|---|
| | $T_c^{zero}$ (K) | $\gamma_s$ |
| RE1 | 92.7 | 8.5 |
| RE2 | 91.9 | 9.5 |
| RE3 | 89.0 | 9 |
| RE4 | 89.9 | 7 |
| RE5 | 88.8 | 6.5 |
| RE6 | 89.8 | 8.5 |
| RE7 | 89.4 | 7.5 |

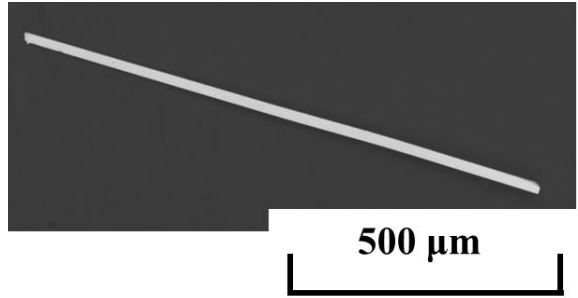

**Figure 1.**

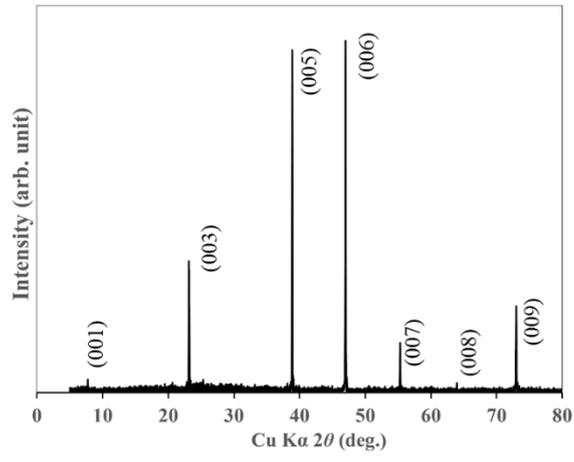

**Figure 2.**

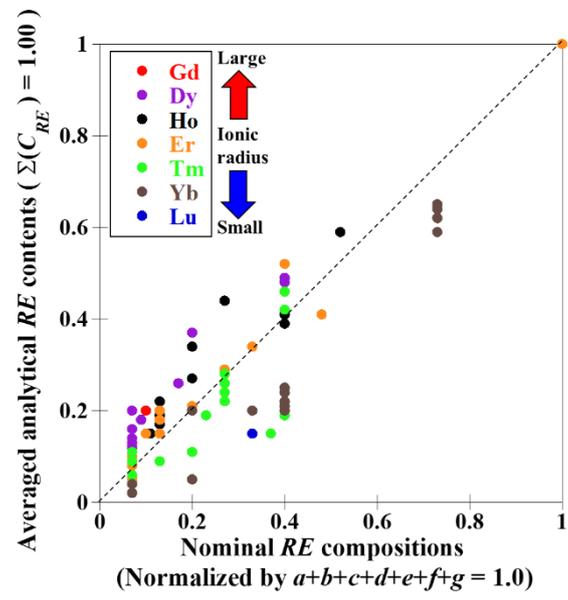

**Figure 3.**

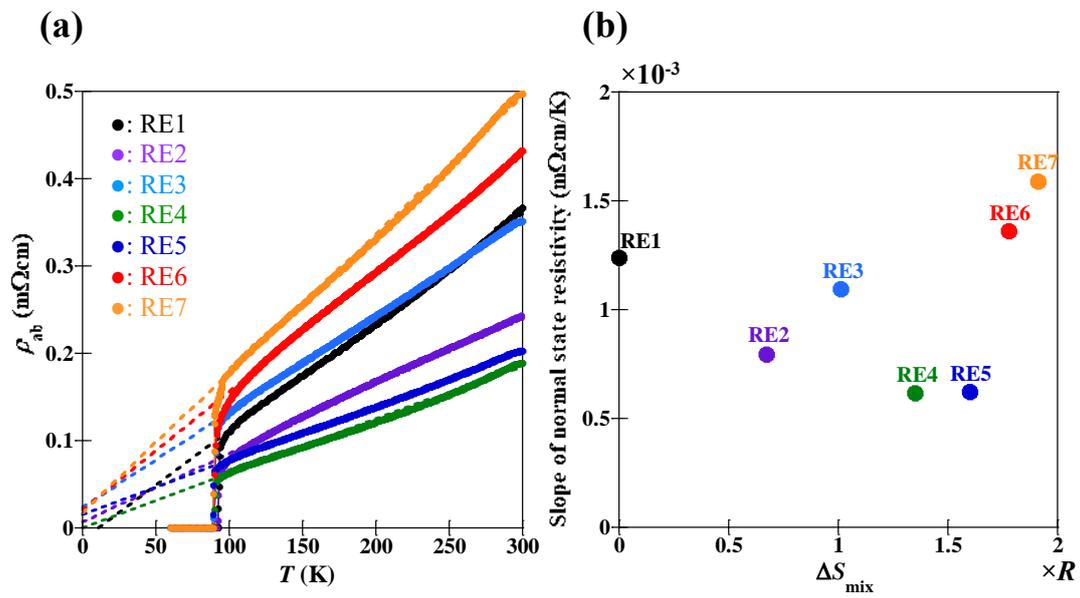

**Figure 4.**

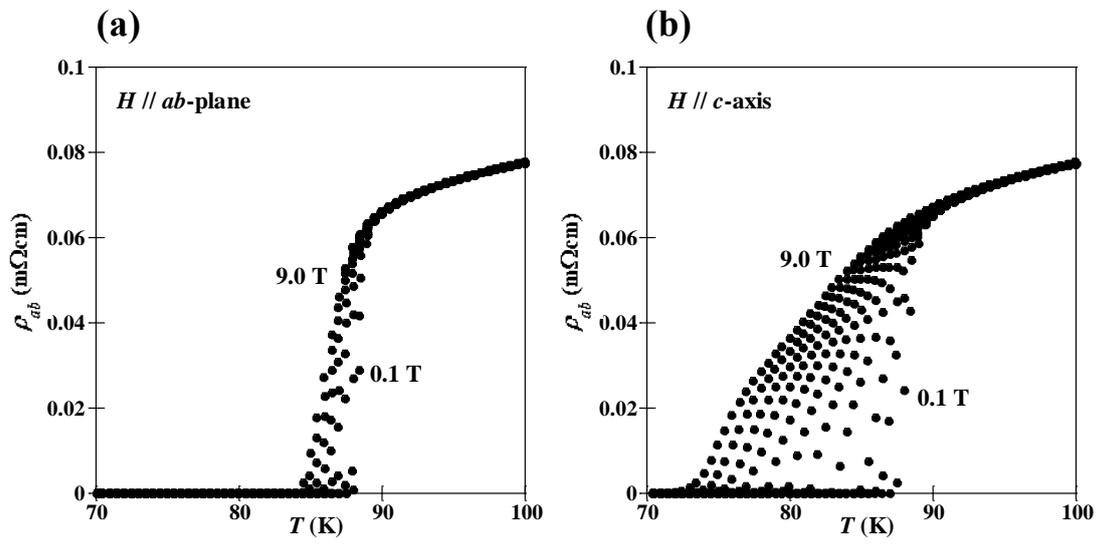

**Figure 5.**

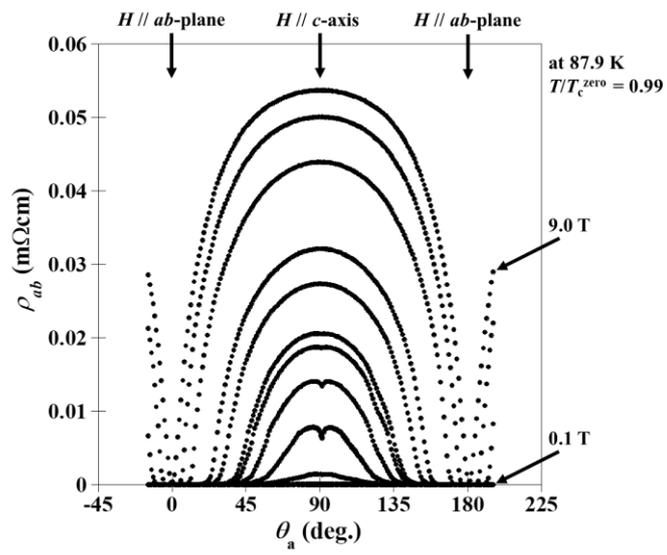

**Figure 6.**

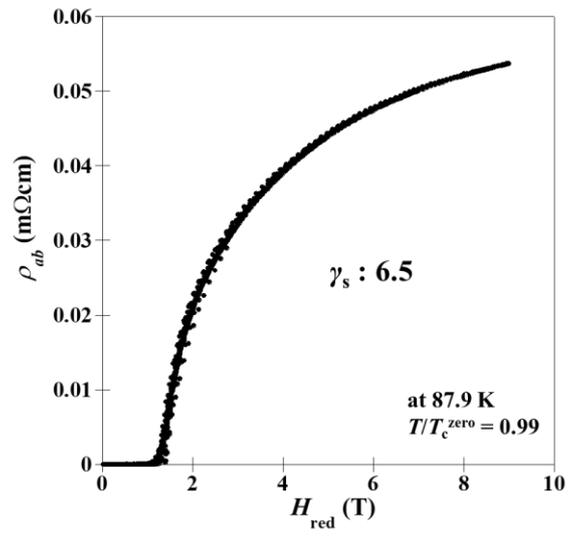

**Figure 7.**

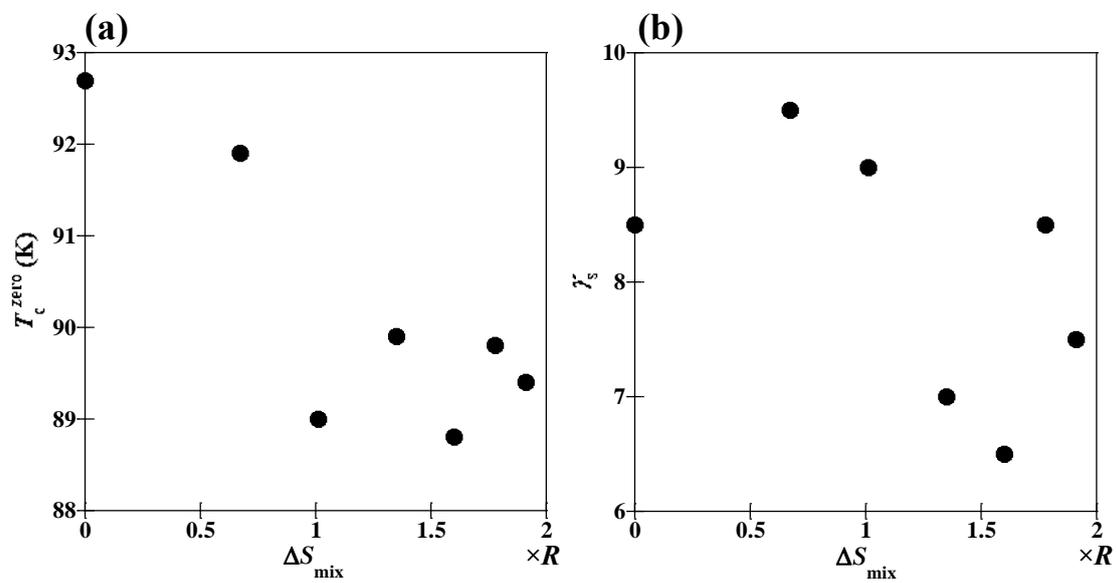

Figure 8.